\documentclass[preprint,proceedings]{rmaa}
\usepackage{paralist}

\usepackage{psfrag,color}

\SetYear{2003}
\SetConfTitle{Winds, Bubbles and Explosions}

\title{High Resolution X-ray Observations of Pulsar Winds}

\author{
  Bryan M. Gaensler\altaffilmark{1} }

\altaffiltext{1}{Harvard-Smithsonian Center for Astrophysics,
Cambridge MA 02138, USA.}

\shortauthor{Gaensler}
\shorttitle{X-ray Observations of Pulsar Winds}

\suppressfulladdresses
\listofauthors{B. M. Gaensler}
\indexauthor{Gaensler, B. M.}

\abstract{Young pulsars spin incredibly quickly, but are also slowing
down at a very rapid rate. This process carries away enormous amounts
of energy from the star in the form of a relativistic wind.  Through
the high resolution now offered by the {\em Chandra X-ray Observatory},
a flood of new observations of pulsars and their winds is demonstrating
the diversity and complexity with which pulsars interact with their
surroundings.  This paper reviews our basic understanding of pulsar
winds, and briefly summarizes some of the new insights provided by
{\em Chandra}\ on shock structure, particle acceleration and wind
composition.}

\resumen{}

\addkeyword{ISM: Jets and outflows}
\addkeyword{Stars: Neutron}
\addkeyword{Supernova remnants}
\addkeyword{X-rays: ISM}

\begin{document}
% Typeset article header
\maketitle

\section{Introduction}
\label{sec:intro}

There is an enormous amount of kinetic energy associated with the rapid
rotation of a young pulsar. For example, a newborn neutron star with an
initial period of 5~ms has a rotational kinetic energy of
$\sim10^{51}$~erg, comparable to that released in the associated
supernova explosion.  Pulsar timing observations clearly demonstrate
that all known isolated pulsars are slowing down, indicating that this
huge reservoir of kinetic energy is being steadily dissipated. The power
associated with this loss of rotational energy (the ``spin-down
luminosity'') is $\dot{E} =
\frac{d}{dt}\left(\frac{1}{2}I\omega^2\right) = 4\pi^2I\dot{P}/P^3$,
where $P = 2\pi/\omega$ is the pulsar spin-period, and $I$ is the
star's moment of inertia (usually assumed to have the value $I \equiv
10^{45}$~g~cm$^2$). For the observed range of $P$ and
$\dot{P}$, we find values up to $\dot{E} \sim10^{39}$~erg~s$^{-1}$,
implying that some pulsars deposit their energy into their
environment at a rate comparable to the bolometric luminosity of the
most massive stars. It is thus reasonable to ask: where does all this
energy go?

In most cases the luminosity of the pulsations themselves is
negligible. It is rather thought that most of a pulsar's spin-down
energy goes into a relativistic wind, populated with electrons,
positrons, and possibly ions as well.  If this wind can be confined,
synchrotron emission will be generated, producing an observable {\em
pulsar wind nebula}\ (PWN).

By far the most well-studied PWN is the Crab Nebula.  The central
pulsar in the Crab has $\dot{E} = 4\times10^{38}$~erg~s$^{-1}$; one
can independently infer that $\sim10^{38}$~erg~s$^{-1}$ must be supplied to the
surrounding nebula to maintain the observed distribution of magnetic
fields and particles. We can conclude that the bulk of a pulsar's
spin-down is deposited into its PWN, and that a PWN thus acts as a
useful probe of a pulsar's energy loss and interaction with its
surroundings.

%HERE WE BRIEFLY REVIEW...

\section{Pulsar Wind Nebulae}

A simple model of pulsar wind nebulae allows one to understand their
basic energetics and structure. Near the pulsar, wind particles flow
freely outwards with zero pitch angle. While this region is not
directly observable, we can infer from theoretical arguments that in
this region the wind has a composition $\sigma \gg 1$ (e.g.\ Coroniti
1990), where $\sigma$ is defined as the ratio of electromagnetic energy
to particle energy in the wind.

At some point from the pulsar the wind will be confined by external
pressure. In this region a termination shock is formed, in which
particles are accelerated and synchrotron radiation is produced.  At
least in the case of the Crab Nebula, observations and modeling of the
PWN emission allow us to determine $\sigma \approx 0.003$ at the
termination shock (Kennel \& Coroniti 1984;
Emmering \& Chevalier 1987), corresponding to a weakly magnetized wind.
It is not clear where or how the wind makes an apparent transition from
$\sigma \gg 1$ to $\sigma \ll 1$, a problem referred to as the
``$\sigma$ paradox'' (Melatos 2002 and references therein).  
Downstream of the termination shock, the wind
decelerates and the pitch angles are randomized; synchrotron emission
is subsequently generated, producing an extended PWN.

Synchrotron emission is an intrinsically broadband process. However, of
particular interest in the case of PWNe is emission in the X-ray band,
because the synchrotron lifetimes of X-ray-emitting electrons are
comparatively short (typically 1--10~years for typical PWN magnetic
fields). This means that X-ray observations of PWNe directly trace the
current conditions in the nebula and near the pulsar. However, these
short lifetimes also implies that the emitting particles generally are not
able to travel far from the pulsar, resulting in a comparatively small
angular extent for the nebular emission. In practice, this means that
previous X-ray images of PWNe, typically with a spatial resolution of
an arcmin or worse, showed interesting morphologies but were unable to
clearly resolve any nebular structures.  Furthermore, observations in
the soft part of the X-ray band suffer from significant photoelectric
absorption, limiting the sensitivity of the resulting images.

Due to these limitations, X-ray studies of pulsars and their nebulae
has long been a frustrating case of ``blobology''.  It is only with the
launch in 1999 of the {\em Chandra X-ray Observatory}\ that major new
advances in this field are finally now being made.  The main reason for
this is {\em Chandra}'s superb subarcsecond resolution, which allows us
to see structures very close to the central pulsar. The {\em Chandra}\
CCDs also have moderate spectral resolution, sufficient for 
spatially resolving the emitting particle distribution in 
these sources.  Finally, {\em
Chandra}'s coverage of the harder parts of the X-ray band (up to
10~keV) ensures that the sensitivity of the observations are not
limited by absorption.

\section{The Crab Nebula} % and Vela Pulsar}

\begin{figure}[!t]
  \includegraphics[width=\columnwidth]{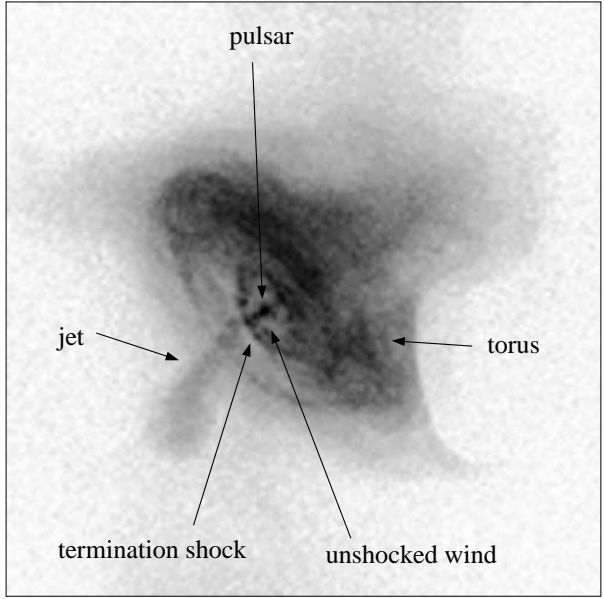}
  \caption{{\em Chandra}\ image of the Crab Nebula (Weisskopf et al.\ 2002).
The central pulsar has an age of 950~yrs, a spin-down
luminosity $\dot{E} = 4\times10^{38}$~erg~s$^{-1}$ and is
at a distance of 2~kpc. The image is $4'\times4'$ in size.}
  \label{fig:crab}
\end{figure}

Not surprisingly, the Crab Nebula was the first PWN looked at by {\em
Chandra}. The resulting image (Figure~\ref{fig:crab}; Weisskopf et al.\ 2000;
Hester et al.\ 2002) 
is nothing short of spectacular,
showing a variety of features not apparent from earlier images. 
These include the central pulsar, a
region of reduced emission around the pulsar corresponding to the
unshocked wind zone, a clear turn-on of emission beyond this zone
representing the termination shock, a broad torus comprising the bulk
of the nebular emission, and a fainter ``jet'' aligned with the axis of
the torus.

The elliptical morphology of the termination shock demonstrates
that the relativistic outflow from the Crab pulsar is not
isotropic, but rather is focused into an equatorial flow. Within
this interpretation, it seems likely that the jet of emission is
directed along the pulsar spin-axis, which is the only fixed
symmetry axis associated with the system. 

%Interestingly, the jet aligns (in projection) with the proper motion
%vector measured for the pulsar.
%
%The Vela pulsar is much older and less energetic than the Crab, but at
%a distance of only $\sim250$~pc it provides a very detailed view of a
%potentially more typical PWN.  The nebula, shown in Fig~XX, is $\sim10$
%times smaller than the Crab, but similarly shows a morphology involving
%an equatorial ring or torus, with a jet-like feature lying along the
%symmetry axis. Again the jet aligns with the pulsar proper motion,
%suggesting a possible general alignment between the spin-axis and the
%velocity vector in the neutron star population (REF).  If such an
%alignment is borne out by a observations of a larger sample, it would
%argue that the angular and linear momenta of a neutron star both have
%their origins in asymmetries in the supernova explosion, as occurs in
%the model involving multiple off-center thrusts proposed by Spruit \&
%Phinney (1998).

\section{Pulsar B1509--58}

PSR~B1509--58 is a young and energetic radio, X-ray and $\gamma$-ray
pulsar. Previous observations have shown this source to be 
surrounded by an
elongated X-ray PWN, with thermal X-rays from the associated
supernova remnant G320.4--1.2 immediately to the north. {\em ROSAT}\
data have revealed a number of interesting features in this source: there is
the suggestion of a collimated outflow along the pulsar spin-axis
like that seen in the Crab Nebula, a possible compact disc
of emission immediately surrounding the pulsar, and
evidence for a ``torus plus jets'' morphology resembling that of 
the Crab (Greiveldinger et al.\ 1995;
Brazier \& Becker 1997). 

\begin{figure}[!t]
  \includegraphics[width=\columnwidth]{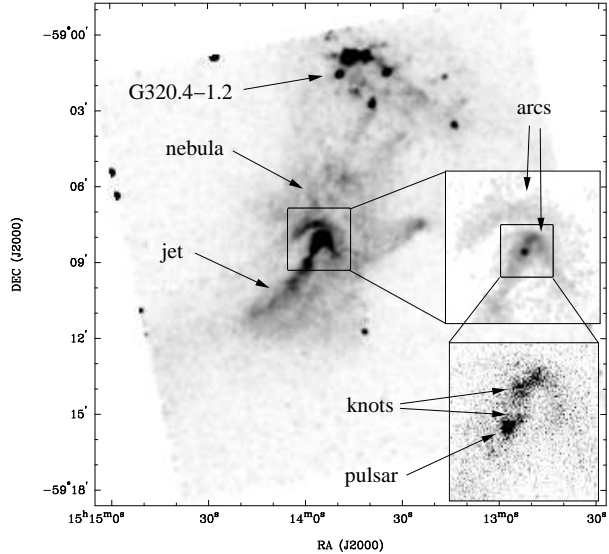}
  \caption{{\em Chandra}\ image of PSR~B1509--58 and its
PWN (Gaensler et al.\ 2002). 
The central pulsar has an age of 1700~yrs, a spin-down
luminosity $\dot{E} = 2\times10{37}$~erg~s$^{-1}$ and is
at a distance of 5~kpc. The two sub-panels show the central
region at successively higher resolution.}
\label{fig:1509}
\end{figure}

We have carried out {\em Chandra}\ observations of this source to
follow up on these possibilities, as reported in detail by Gaensler et
al.\ (2002). The resulting image, presented in Figure~\ref{fig:1509},
shows the pulsar itself, surrounded by a diffuse elongated nebula, with
a clear jet-like structure lying to the south of the pulsar along the
main symmetry axis of the nebula. There is also an arc of emission
immediately to the north of the pulsar, bisected by this symmetry axis.
At higher resolution, a second arc is seen nestling inside the main
arc.  Even closer to the pulsar is a collection of 3--4 knots of X-ray
emission.

The {\em Chandra}\ data demonstrate that there is a clear symmetry axis
associated with this system, defined on scales ranging from $10''$ up
to $10'$. As with the Crab Nebula, it is most likely that this axis
represents the pulsar spin axis. The collimated, curved jet-like
feature also resembles that seen for the Crab. This feature has a
harder X-ray spectrum than the rest of the PWN.  If this hard spectrum
results from a lack of synchrotron cooling compared to surrounding
emission, the more rapid rate at which particles in this region must be
replenished implies a flow velocity along this feature of $>0.2c$
(Gaensler et al.\ 2002). Thus it appears that this structure is indeed a true
jet, which is somehow being accelerated and collimated by the pulsar.

The morphology of the arcs seen to the north of the pulsar suggest that
they are circular rings in an inclined equatorial plane.
Gaensler et al.\ (2002) have interpreted these arcs in the
context of the PWN model of Gallant \& Arons (1994), who argue that
ions in the wind undergo magnetic reflection in the termination shock
zone, and that the electrons are consequently compressed at the ion
turning points. This produces synchrotron ``wisps'' where the electrons
are compressed, whose predicted spacing ($\sim$~1.8 to 1) match that of
the two arcs seen here. The particle flux through these features allows
us to calculate a magnetization parameter at the termination shock of
$\sigma \approx 0.005$, comparable to that previously determined for
the Crab Nebula.

Within this interpretation, the knots seen very close to the pulsar
then represent emission from the {\em unshocked}\ wind,
upstream of the termination shock. While the process through which
these features are produced is as yet uncertain, we assume that
their observed width represents a Larmor orbit of the emitting
particles. The consequent magnetic field allows us to infer that
$\sigma < 0.003$ at a separation from the pulsar of $\la 0.1$~pc (see
Gaensler et al.\ 2002 for details). Thus, if there is truly a transition
in the wind from $\sigma \gg 1$ to $\sigma \ll 1$, it occurs much
closer to the pulsar than can be resolved even with {\em Chandra}.

\section{Conclusions}

The detailed {\em Chandra}\ data-sets now being obtained on these and
other PWNe conclusively demonstrate that X-ray observations at high
resolution give us our best insight yet into the processes through
which pulsars lose their rotational energy.  Fundamental similarities
are now emerging amongst the wind properties of the pulsar population,
but this new level of detail also makes clear that there are important
(and yet to be understood) differences.

\acknowledgements

I thank 
Jon Arons for his enthusiasm
and input in interpreting the PWN surrounding PSR~B1509--58. This work
has been supported by NASA through grants SAO-GO0-1134X and
NAG5-11376.

\end{document}